# Rank Based Clustering For Document Retrieval From Biomedical Databases

Jayanthi Manicassamy
Department Of Computer Science,
Pondicherry University,
Kalapet, India.
jmanic2@yahoo.com

P. Dhavachelvan
Department Of Computer Science,
Pondicherry University,
Kalapet, India.

*Abstract*— Now a day's, search engines are been most widely used for extracting information's from various resources throughout the world. Where, majority of searches lies in the field of biomedical for retrieving related documents from various biomedical databases. Currently search engines lacks in document clustering and representing relativeness level of documents extracted from the databases. In order to overcome these pitfalls a text based search engine have been developed for retrieving documents from Medline and PubMed biomedical databases. The search engine has incorporated page ranking bases clustering concept which automatically represents relativeness on clustering bases. Apart from this graph tree construction is made for representing the level of relatedness of the documents that are networked together. This advance functionality incorporation for biomedical document based search engine found to provide better results in reviewing related documents based on relativeness.

## I. INTRODUCTION

At present, there found to be a tremendous movement towards development of technologies that are been establishing in each and every area involving bioinformatics. Previously area of bioinformatics found to be lacking but now, this area found to have a tremendous development comparatively to other areas [1-5]. Today, rapidly increasing volume of publications in the biomedical, finding related work is an ever more difficult challenge. General solutions to the document search problems are difficult because biomedical science is very diverse for the articles most relevant to the readers the relevancy may vary. Relevance is a well-established technique to improve performance in information retrieval. Hierarchical classification has received growing attention in the biomedical field in recent years which is used for various biological data's related classification specifically in the field of text mining [10-15]. In the study made it has been found that hierarchical classification found to have better results of utilization which provoked in the development of a text based search engine which have been narrated in this article.

The developed text based search engine is capable of retrieving biomedical documents from biomedical databases Medline and PubMed that are clustered based on the relativeness of the document to the user search. Clustering based on page ranking which represents the level of relativeness for the retrieved clustered documents. Document retrieval is based on the occurrence of biomedical terminologies and keywords based on the user search text. Considerations taken for document retrieval are mainly considered here are both on positional and relationship apart from the other criteria's has been described in section II. The developed tool have been evaluated which have been narrated in section III in results and discussions.

## II. SYSTEM DESCRIPTION

### A. Implementation

In this section system implementations along with architecture have been discussed. The system developed is a text based search engine which is capable of extracting the documents from Medline and PubMed databases. Figure 1 shows the system architecture in which rectangle represents tasks. Dotted lines represent sub tasks links and streamed line represents major tasks involved for retrieving documents and rounded corner box represents activities. The only position in the system where human intervention is required is entering the text to be searched for documents retrieving in this system.

For the developed system search text is inputted by the user for which related and relevant documents have to be retrieved. The following are major tasks involved in processing the texts and retrieving the documents.

### 1) Analyzer

This is the first major task that is carried out each and every time when texts are entered for document retrieving. Keywords, biomedical terminologies are extracted from the texts that are to be searched for documents. From the extracted keywords and terminologies related or equivalent alternative words are extracted through wordnet and stored in a relation list temporarily through relation setter.

**Figure 1: System Architecture**

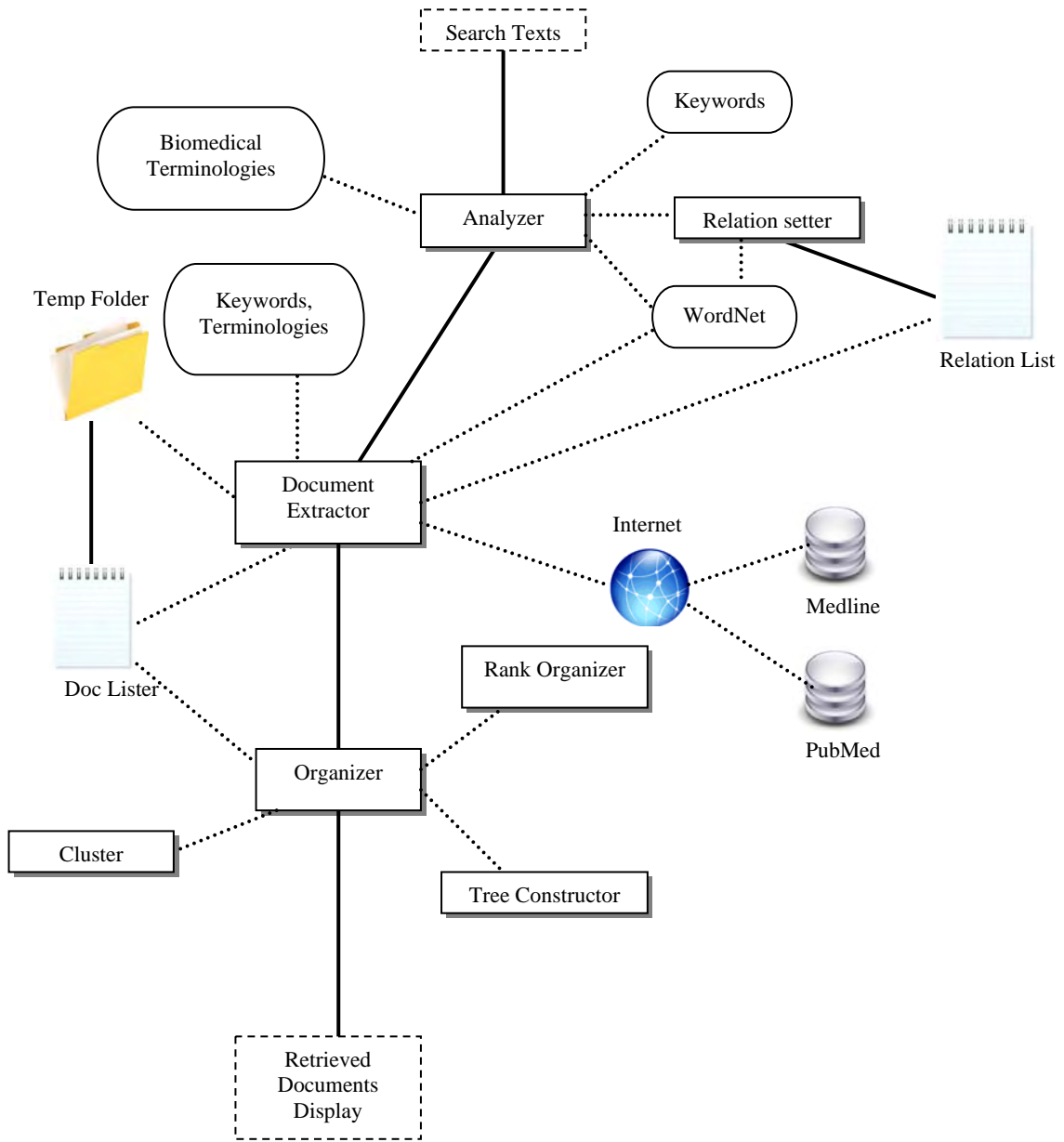

*2) Document Extractor*

Based on the relation list related and relevant documents are extracted from Medline and PubMed biomedical databases based on keywords and terminologies based on the comparison made with that of relation list. The extracted documents are stored in temp folder. The document with related keywords based on which document have been extracted and stored in temp folder is listed in doc lister and stored in the same temp folder for carrying out the next major task.

*3) Organizer*

The task of this organizer is to rank the documents based on rank score through rank organizer by utilizing doc lister. The next task of this organizer is to cluster the documents based on rank score with represents the level of relativeness of the text search made by the system which is then constructed in graph tree through tree constructor.

**Figure 2: Model Graph Tree Structure**

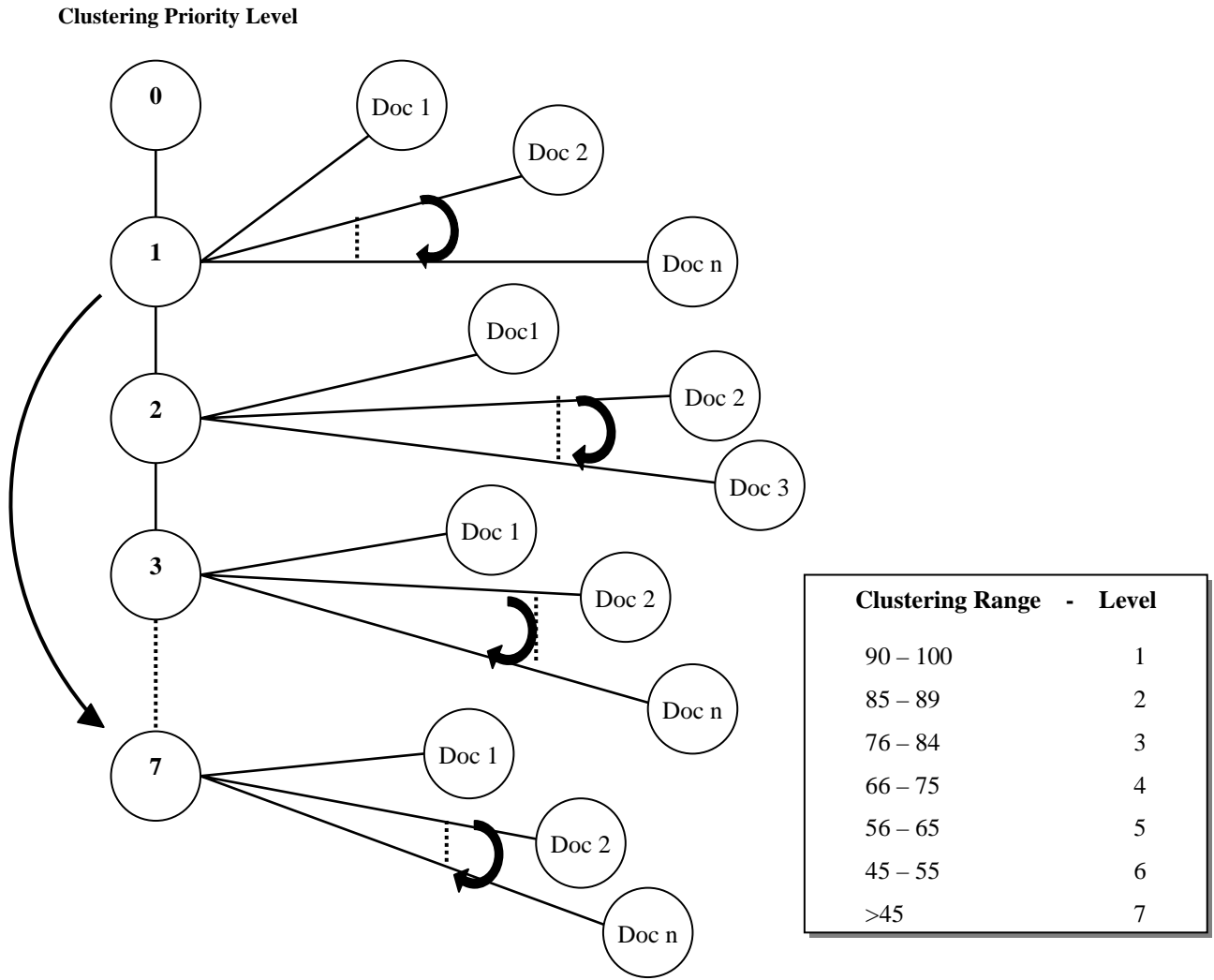

The document extracted are visualized in graph tree with parent nodes representing level of relativeness and sub node of the parent nodes represents document placed in hierarchical level which could be review by clicking each nodes.

B. *Working Methodology*

In this section processing methodology of the system have been narrated in detail as follows

1) *Analyzer*

Inputted text processing is initialized at this step which takes in the search text which is analyzed for keywords,

biomedical terminologies and sets relational list by using relation setter.

2) *Relation Setter*

The activity of this relation setter is to identify keywords from the texts that are to be searched for which biomedical terminologies are spitted. Both keywords and terminologies are temporally stored. By using both keywords and terminologies as base usage WordNet all possible alternative related words are extracted and listed in relation list based on the relativeness. The priority of positioning the keywords and terminology in the list is first exact terminologies and keywords followed by related terminologies and keywords.

3) *Document Extractor*

Documents are extracted from Medline and PubMed databases based on extracting keywords and terminologies

Jayanthi Manicassamy et al /International Journal on Computer Science and Engineering Vol.1(2), 2009, 111-115from the documents and making a comparison with that of relation list. If match found then the document are been listed with matched keyword in doc list. Both doc list and the matched documents are stored in a temp folder. Doc list consists of relationship of the keywords to the document.

*4) Organizer*

Doc lister is used inorder to carry out further activities in documents displaying for the search made by the organizer. Based on keywords and terminologies rank score is through rank organizer rank score of extracted documents are evaluated based on keywords and terminologies. For rank score evaluation each direct match keyword ($d_k$) and direct match terminologies ($d_t$) value 1. Where as for indirect direct match keyword ($id_k$) value 0.5 and indirect match biomedical terminologies ($id_t$) value 0.8.

$$DS_i = \frac{\sum d_k + \sum d_t + \sum id_k + \sum id_t}{\sum K_i + \sum T_i} + \sum Kw_i \quad (1)$$

$$CL_i = \frac{\sum d_k + \sum d_t + \sum id_k + \sum id_t}{\sum K_i + \sum T_i} \times 100 \quad (2)$$

$$d = \frac{\sum d_k + \sum d_t}{\sum K_i + \sum T_i} \times 100 \quad (3)$$

$$id = \frac{\sum id_k + \sum id_t}{\sum K_i + \sum T_i} \times 100 \quad (4)$$

Here, $DS_i$ represent document rank score, $K_i$ are keywords excluding biomedical terminologies and $T_i$ are biomedical terminologies found in user entered search text. $Kw_i$ is total keywords and biomedical terminologies match the text search found from the document. $CL_i$ represents Clustering Level of relativeness for the texts search made for document retrieval. d denotes level of direct keywords and terminologies matches the present document compared with that of the search documents. id denotes level of indirect keywords and terminologies matches the present document compared with that of the search documents. Clustering of documents and positioning the documents in the cluster depends on $CL_i$, d and id. If d found to higher compare to id than 0.2 is add to $DS_i$.

Graph Tree construction id made based on the clustering and ranking of the documents. Tree construction groups documents and represented in a tree structure based on relativeness of the document that have been represented as in figure 2. In the figure 1 to 7 represents cluster level of relativeness which are parent nodes. While child nodes are documents extracted from the databases. The approach is a top down approach where cluster level 1 is high where the level increases the level of relativeness of the documents to the search text decreases. doc 1 to doc n represents the position and rank of the document in that cluster. The document extracted are visualized in graph tree with parent nodes representing level of relativeness and sub node of the parent nodes represents document placed in hierarchical level where top position found to high comparatively to the next levels. Documents could be viewed by clicking on the respective child nodes.

III. EVALUATION AND DISCUSSION

In this section evaluation on the developed system has been carried out for which the performance doesn't lie in one working step of the system. Where as each and every process are responsible for system performance for which the major performance lies in the analysis part. For various evaluations carried out on the system the performance found to be good. The only place where human intervention is required is entering the texts for search in retrieving documents form Medline and PubMed databases. The evaluation of the system found to have precision of 87% and recall found to be 89% which is found to be high of which their found to be less variation in performance.

It is well known that, there exists lots of information needs related to biomedical area which vary in functionality where most of them fall online. The main novelty lies in clustering the documents based on relativeness and representing in a graph tree structure for displaying documents. Documents are represented as child nodes and reviewing each document is by clicking the each child node. The system utilization could be very beneficial for the community in the field of biomedical for reviewing document based on the relativeness of the search text formulated by users. This would be very helpful in terms of time and effort for reviewing only the required highly relative documents.

IV. CONCLUSION

Currently developed text based search system found to be highly significant in biomedical domain for document retrieval. This system shows an improvement over the existing systems with better results which offer new information representation capabilities with different techniques like clustering based on relativeness of the document to the search texts. Apart from this, making users to identify level of relativeness of the clusters and documents ranks in a graph tree structure. From various evaluations carried out the performance of the system found to be good comparatively to other systems in biomedical domain.

REFERENCES
[1] Jayanthi Manicassamy and P. Dhavachelvan, "Metrics based performance control over text mining tools in bioinformatics", ACM Portal, Pages 171-176, January 2009.
[2] Jayanthi Manicassamy and P. Dhavachelvan, "Based Accuracy Perpetuation for Bioinformatics Sequence Analysis Tools", International Journal of Recent Trends in Engineering (IJRTE) - Finland, pp 550-555, May 2009.

[3] Marta Sabou, Chris Wroe, Carole Goble, Gilad Mishne, "Learning domain ontologies for web service descriptions: An experiment in bioinformatics", citeseer, May 2005.

[4] Richard Tzong-Han Tsai, Shih-Hung Wu, Wen-Chi Chou, Yu-Chun Lin, Ding He, Jieh Hsiang, Ting-Yi Sung and Wen-Lian Hsu, "Various criteria in the evaluation of biomedical named entity recognition", PubMed, pp7-92, February 2006.

[5] Jung-jae Kim, Piotr Pezik and Dietrich Rebholz-Schuhmann, "MedEvi: Retrieving textual evidence of relations between biomedical concepts from Medline", ACM portal, pp1410–1412, March, 2008.

[6] Arek Gladki, Pawel Siedlecki, Szymon Kaczanowski and Piotr Zielenkiewicz, "e-LiSe—an online tool for finding needles in the '(Medline) haystack'", ACM Portal, pp1115–1117, March 2008.

[7] Mario Falchi and Christian Fuchsberger, "Jenti: an efficient tool for mining complex inbred genealogies", ACM Portal, pp724–726, January 2008.

[8] Daraselia N, Yuryev A, Egorov S, Mazo I, Ispolatov I, "Automatic extraction of gene ontology annotation and its correlation with clusters in protein networks", BMC Bioinformatics 2007, Vol 8 (1), pp-243.

[9] Tsoumakas G, Katakis I: Multi-Label Classification, "An Overview. International Journal of Data Warehousing and Mining", 2007, Vol 3(3), pp 1-13.

[10] Barutcuoglu Z, Schapire RE, Troyanskaya OG, "Hierarchical multilabel prediction of gene function", Bioinformatics, 2006, Vol 22(7), pp 830-836.

[11] Cai L, Hofmann T, "Hierarchical document categorization with support vector machines", ACM 13th Conference on Information Management, 2004.

[12] Dumais ST, Chen H, "Hierarchical classification of web content", ACM Special Interest Group on Information Retrieval (SIGIR), 2000, pp 256-263.

[13] Rousu J, Saunders C, Shawe-Taylor J, "Kernel-based learning of hierarchical multilabel classification models", Journal of Machine Learning Research, 2006, Vol 7, pp 1601-1626.

[14] Verspoor K, Cohn J, Mniszewski S, Joslyn C, "A categorization approach to automated ontological function annotation", 2006, Vol 15(6), pp 1544-1549.

[15] Wolstencroft K, Lord P, Tabernero L, Brass A, Stevens R, "Protein classification using ontology classification", Bioinformatics 2006, Vol 22(14), pp 530-538.